\documentstyle[twocolumn,aps]{revtex}
\begin{document}
\draft
\title{Thermal stability of coupled ferromagnetic and superparamagnetic particles}
\author{Vladimir L. Safonov \thanks{%
E-mail: vsafonov@ucsd.edu} and H. Neal Bertram}
\address{Center for Magnetic Recording Research, University of California - San Diego,%
\\
9500 Gilman Dr., La Jolla, CA 92093-0401, U.S.A.}
\date{\today}
\maketitle

\begin{abstract}
We consider a single-domain ferromagnetic particle with uniaxial anisotropy
coupled to a single-domain soft ferromagnetic particle (superparamagnetic
particle). The problem of thermally agitated magnetization reversal in this
case can be reduced to the random magnetization dynamics of the first
particle with an effectively larger anisotropy field. The magnetic external
field is also altered in a manner that depends on the sign of the coupling
and can be either enhanced or suppressed.
\end{abstract}


\section{Introduction}

Thermal stability is of great importance in ultra-high-density magnetic
recording. The principal element of magnetic memory is a single-domain
ferromagnetic particle. The magnetic moment of this particle is agitated by
thermal fluctuations and can reverse to the opposite direction. In the
absence of inter-particle interactions the magnetization reversal time is
determined by the N\'{e}el-Arrhenius exponent with a definite energy barrier
(e.g., \cite{brown},\cite{weller},\cite{hans}). The problem for the mean
magnetization reversal time has been rigorously solved in the framework of a
new theoretical approach, in which the coherent rotation of the
magnetization is analyzed in terms of the motion of a damped nonlinear
oscillator subject to random Gaussian forces (thermal agitation) \cite
{safonov},\cite{safbert}, \cite{bertsaf},\cite{safbertbook},\cite{wbs}.

Recently it has been experimentally demonstrated that the thermal stability
of magnetic memory can increase if the ferromagnetic particles from the
recording layer are antiferromagnetically coupled to an auxiliary layer of
smaller ferromagnetic particles (AFC media) \cite{abarra},\cite{fullerton}.
Simple theoretical explanations of this phenomenon are given, for example,
in Refs.\cite{scabes},\cite{richter}. The basic model consists of two
single-domain hard particles (recording and stabilizing) coupled by an
antiferromagnetic exchange interaction.

Here we analyze the possibilities of the two-particle model to enhance
thermal stability of recording media. We consider the thermally agitated
magnetization reversal, when a hard ferromagnetic particle interacts with a
soft ferromagnetic (superparamagnetic) particle. We use the new theoretical
approach for coupled damped nonlinear oscillators subject to random Gaussian
forces introduced by thermal agitation.

\section{Model}

Let us consider a ferromagnetic particle with magnetization ${\bf M}_{1}$($|%
{\bf M}_{1}|=M_{s,1}$) and volume $V_{1}$ and a superparamagnetic particle
with ${\bf M}_{2}$ ($|{\bf M}_{2}|=M_{s,2}$) and $V_{2}$, respectively
(Fig.1). To describe the coherent motion of the particle magnetizations we
introduce the classical spins ${\bf S}_{j}={\bf M}_{j}V_{j}/\hbar \gamma $ ($%
j=1,2$), where $\gamma $ is the gyromagnetic ratio and $\hbar $ is Planck's
constant.

The energy of the combined system includes uniaxial anisotropy of the
ferromagnetic particle along $z$ axis, Zeeman energy, exchange and
magnetostatic interparticle interactions:

\begin{eqnarray}
{\cal E} &=&-K_{u}V_{1}\left( \frac{S_{1}^{z}}{S_{1}}\right) ^{2}-\hbar
\gamma {\bf H}_{0}\cdot ({\bf S}_{1}+{\bf S}_{2})+J{\bf S}_{1}\cdot {\bf S}%
_{2}  \nonumber \\
&&+(\hbar \gamma )^{2}\left[ \frac{{\bf S}_{1}\cdot {\bf S}_{2}}{r^{3}}-3%
\frac{({\bf S}_{1}\cdot {\bf r)(S}_{2}\cdot {\bf r)}}{r^{5}}\right] .
\label{energy}
\end{eqnarray}
Here $K_{u}$ is the uniaxial anisotropy constant, ${\bf H}_{0}$ is the
external magnetic field and $J$ is the exchange integral ($J>0$ corresponds
to antiferromagnetic and $J<0$ to the ferromagnetic coupling, respectively).
The last term in (\ref{energy}) describes the magnetostatic interactions
(dipole-dipole approximation), where ${\bf r}$ is the vector between
particle centers.

For simplicity we shall confine our analysis to two particle configurations
applicable for the longitudinal (Fig.1a) and perpendicular (Fig.1b)
recording media with magnetic fields parallel to each uniaxial anisotropy
axis. Thus Eq.(\ref{energy}) becomes

\begin{eqnarray}
{\cal E} &=&-K_{u}V_{1}\left( \frac{S_{1}^{z}}{S_{1}}\right) ^{2}-\hbar
\gamma H_{0}(S_{1}^{z}+S_{2}^{z})  \label{energyapp} \\
&&+\hbar \gamma (H_{int}S_{1}^{z}S_{2}^{z}+H_{\perp
,x}S_{1}^{x}S_{2}^{x}+H_{\perp ,y}S_{1}^{y}S_{2}^{y}).  \nonumber
\end{eqnarray}
Here $H_{\perp ,y}=J/\hbar \gamma +\hbar \gamma /r^{3}$. For configurations
shown in Fig.1 the interaction fields $H_{int}$ and $H_{\perp ,x}$\ are
defined by 
\begin{eqnarray}
H_{int}^{(a)} &=&\frac{J}{\hbar \gamma }+\frac{\hbar \gamma }{r^{3}},
\label{Hinta} \\
H_{\perp ,x}^{(a)} &=&\frac{J}{\hbar \gamma }-2\frac{\hbar \gamma }{r^{3}} 
\nonumber
\end{eqnarray}
for Fig.1a and 
\begin{eqnarray}
H_{int}^{(b)} &=&\frac{J}{\hbar \gamma }-2\frac{\hbar \gamma }{r^{3}},
\label{Hintb} \\
H_{\perp ,x}^{(b)} &=&\frac{J}{\hbar \gamma }+\frac{\hbar \gamma }{r^{3}} 
\nonumber
\end{eqnarray}
for Fig.1b. 

\section{Dynamic equations}

In order to describe spin (magnetization) dynamics, we introduce complex
variables $a_{j\ }^{\ast }$ and $a_{j}$\ (classical analog of creation and
annihilation operators) of nonlinear oscillators defined by the
Holstein-Primakoff transformation \cite{hopri}:

\begin{eqnarray}
S_{j}^{z} &=&S_{j}-N_{j},\quad N_{j}=a_{j}^{\ast }a_{j},  \nonumber \\
S_{j}^{x} &=&\frac{a_{j}+a_{j}^{\ast }}{2}\sqrt{2S_{j}-N_{j}},  \nonumber \\
S_{j}^{y} &=&\frac{a_{j}-a_{j}^{\ast }}{2i}\sqrt{2S_{j}-N_{j}}.  \label{h-p}
\end{eqnarray}
This transformation gives a convenient representation of spins as
oscillators in the vicinity of equilibrium ${\bf S}_{1}=(0,0,S_{1})$ and $%
{\bf S}_{2}=(0,0,S_{2})$.

The dynamic equations for $a_{1}$ and $a_{2}$, defined by $%
da_{j}/dt=-i\partial ({\cal E}/\hbar )/\partial a_{j}^{\ast }$, can be
written as

\begin{eqnarray}
\frac{d}{dt}a_{1} &=&-i\gamma H_{eff,1}a_{1}-i\gamma F_{12},  \label{equc} \\
\frac{d}{dt}a_{2} &=&-i\gamma H_{eff,2}a_{2}-i\gamma F_{21}.  \nonumber
\end{eqnarray}
Here

\begin{eqnarray}
H_{eff,1} &=&H_{0}+H_{K}S_{1}^{z}-H_{int}S_{2}^{z},  \label{efffield1} \\
H_{eff,2} &=&H_{0}-H_{int}S_{1}^{z}  \label{efffield2}
\end{eqnarray}
are the effective fields, $H_{K}=2K_{u}/M_{s,1}$ is the anisotropy field and

\begin{equation}
F_{mn}=H_{\perp ,x}\frac{\partial S_{n}^{x}}{\partial a_{n}^{\ast }}%
S_{m}^{x}+H_{\perp ,y}\frac{\partial S_{n}^{y}}{\partial a_{n}^{\ast }}%
S_{m}^{y}.  \label{fastoscill}
\end{equation}

The principal oscillatory motion of $a_{1}(t)$ and $a_{2}(t)$ are determined
by their effective fields: $a_{1}\propto \exp (-i\gamma H_{eff,1}t)$ and $%
a_{2}\propto \exp (-i\gamma H_{eff,2}t)$. Because, in general, $%
H_{eff,1}\neq H_{eff,2}$, the fast oscillating terms $F_{12}$ and $F_{21}$
in (\ref{equc}) vanish. The magnetization fluctuations of the
superparamagnetic particle occurs at a much faster rate $\sim f_{2}$ than
the reversal rate of the ferromagnetic particle. $f_{2}$ is an attempt
frequency of the second particle (a negligibly small energy barrier is
assumed). Thus, the longitudinal component $S_{1}^{z}$ is the slowest
variable in the system, and the analysis of the coupled system can be
reduced to the stochastic dynamics of the first particle in an averaged
field from the second particle. This field appears as a feedback of the
superparamagnetic particle on the instantaneous ordered state $S_{1}^{z}$ of
the first particle. The corresponding effective energy is equal to

\begin{equation}
{\cal E}_{2}\simeq -\hbar \gamma H_{eff,2}\ S_{2}^{z}.  \label{energy2}
\end{equation}

In this mean field approximation (\ref{energy2}), it follows that for times $%
\gg $ $f_{2}^{-1}$ an averaged thermal spin polarization of the second
(superparamagnetic) particle (e.g. \cite{Aharoni}) is: 
\begin{equation}
\langle S_{2}^{z}\rangle =S_{2}B_{S_{2}}\left( S_{2}\frac{\hbar \gamma
H_{eff,2}}{k_{B}T}\right) ,  \label{polarization}
\end{equation}
where 
\begin{equation}
B_{S}(Sx)=1-\frac{1}{S}\left( \frac{1}{\exp (x)-1}-\frac{2S+1}{\exp
[(2S+1)x]-1}\right)   \label{Brillouin}
\end{equation}
is the Brillouin function. This averaged spin polarization $\langle
S_{2}^{z}\rangle $ induces an effective field (or, response field) back to
the first particle. As a result the effective energy of the first particle
can be written as

\begin{equation}
{\cal E}_{1}\simeq -\frac{\hbar \gamma H_{K}}{2S_{1}}\left( S_{1}^{z}\right)
^{2}-\hbar \gamma \left( H_{0}-H_{int}\langle S_{2}^{z}\rangle \right)
S_{1}^{z}.  \label{energy1}
\end{equation}
From (\ref{Hinta}) and (\ref{Hintb}) one can see that $H_{int}$ is enhanced
for antiferromagnetic ($J>0$) exchange in the longitudinal case (Fig.1a) and
for ferromagnetic exchange in the perpendicular case (Fig.1b).

\section{Approximate solution}

Let us consider a small polarization of the superparamagnetic particle. In
this case Eq.(\ref{polarization}) is simplified to 
\begin{equation}
\langle S_{2}^{z}\rangle \simeq \frac{S_{2}(S_{2}+1)}{3}\frac{\hbar \gamma
H_{eff,2}}{k_{B}T}  \label{smallpolariz}
\end{equation}
and the effective energy (\ref{energy1}) becomes

\begin{equation}
{\cal E}_{1}\simeq -\frac{\hbar \gamma \widetilde{H}_{K}}{2S_{1}}\left(
S_{1}^{z}\right) ^{2}-\hbar \gamma H_{eff,1}S_{1}^{z}.  \label{energy1a}
\end{equation}
Here 
\begin{equation}
\widetilde{H}_{K}=H_{K}\left( 1+2S_{1}\zeta \frac{H_{int}}{H_{K}}\right)
\label{effHK}
\end{equation}
is the effective anisotropy field and

\begin{equation}
H_{eff,1}=H_{0}\left( 1-\zeta \right)  \label{Heff1}
\end{equation}
is the effective external field on the first particle,

\begin{equation}
\zeta =\frac{S_{2}(S_{2}+1)}{3}\frac{\hbar \gamma H_{int}}{k_{B}T}.
\label{zeta}
\end{equation}

From (\ref{effHK}) and (\ref{zeta}) we see that the effective anisotropy
field $\widetilde{H}_{K}$ always increases ($H_{int}\zeta \propto H_{int}^{2}
$)\ for any $H_{int}\neq 0$. On the contrary, the effective external
magnetic field $H_{eff,1}$ (\ref{Heff1}) depends on the sign of $H_{int}$:
it can be both enhanced ($\zeta \propto H_{int}<0$) and suppressed ($\zeta >0
$). The feedback effect from the superparamagnetic particle quadratically
depends on the saturation magnetization and volume of this particle: $\zeta
\propto S_{2}^{2}\propto M_{2}^{2}V_{2}^{2}$. This quadratic dependence also
means that the corresponding feedback effect of two superparamagnetic
particles of volume $V_{2}/2$ is at least two times smaller. Thus it is more
effective to use a single larger superparamagnetic particle than a
collection of smaller ones.

The energy (\ref{energy1a}) is an exact analog to the case of random motion
of one single-domain particle when the external magnetic field is parallel
to the anisotropy axis. Thus, we can use exact analytic formulas \cite
{safbert},\cite{bertsaf},\cite{safbertbook} for the mean first passage time
obtained for this case. It should be noted that this approach makes it
possible to describe experimental data for the reversal field versus pulse
time over 13 orders from nanoseconds to hours \cite{bertsaf}, including both
purely dynamic and thermally assisted reversal.

\section{Results and discussion}

Let us estimate Eqs. (\ref{effHK}) and (\ref{Heff1}). For the ferromagnetic
particle we take the following parameters at room temperature: $%
K_{u}V_{1}/k_{B}T=60$, $H_{K}=5$ kOe and $M_{s,1}=150$ emu/cc. This means $%
V_{1}\simeq 6.62\cdot 10^{-18}$ cm$^{3}$ and $S_{1}\simeq 5.36\cdot 10^{4}$.
Taking $S_{2}\simeq 700$ and $H_{int}=0.3$ Oe, one obtains: $\zeta \simeq
0.022$ and $2S_{1}\zeta H_{int}/H_{K}\simeq 0.14$. Thus, we see that the
effect of the superparamagnetic particle is relatively small in the
effective magnetic field $H_{eff,1}\simeq 0.978H_{0}$ and larger in the
effective anisotropy field $\widetilde{H}_{K}=1.14H_{K}$. For the opposite
sign of interaction (ferromagnetic exchange between particles) $H_{int}=-0.3$
Oe and with the same parameters, one has $H_{eff,1}\simeq 1.022H_{0}$ and $%
\widetilde{H}_{K}=1.14H_{K}$.

In Fig.2 reversal field versus pulse time is shown for the above two cases.
We also show the corresponding curve for one ferromagnetic particle ($%
H_{int}=0$) and Sharrock's (N\'{e}el-Arrhenius) dependence. The thick solid
line and broad dashed curve represent antiferromagnetic and ferromagnetic
coupling, respectively. The slight difference is due to the larger effective
field and hence reduced stability of the ferromagnetic case. In the thermal
region ($H<H_{K}$) there is more curvature than in the N\'{e}el-Arrhenius
approximation. This is because the nonlinear oscillator model contains a
field dependent attempt frequency. Note that formulas (\ref{energy1a})-(\ref
{zeta}) are valid for long time scale $\gg $ $f_{2}^{-1}\sim 10^{-10}-10^{-9}
$ sec. In the dynamic reversal region particles seems to reverse
independently. 

\section*{Acknowledgment}

This work was partly supported by matching funds from the Center for
Magnetic Recording Research at the University of California - San Diego and
CMRR incorporated sponsor accounts.

\bigskip

{\large Figure captions}

Fig.1. Particle configurations for a) longitudinal and b) perpendicular
recording. The larger volume represents the ferromagnetic particle.

Fig.2. Reversal field versus pulse time. 

\end{document}